# High Aspect Pattern Formation by Integration of Micro Inkjetting and Electroless Plating

P. W. Gian[1], Xuechuan Shan[*1], Y. N. Liang[1], B. K. Lok[1], C. W. Lu[1] and B. L. Ooi[2]

[1] Singapore Institute of Manufacturing Technology (SIMTech), Singapore 638075
[2] Electrical & Computer Engineering, National University of Singapore, Singapore 117576

* E-mail: xcshan@simtech.a-star.edu.sg (Xuechuan Shan)

*Abstract* -This paper reports on formation of high aspect micro patterns on low temperature co-fired ceramic (LTCC) substrates by integrating micro inkjetting with electroless plating. Micro inkjetting was realized by using an inkjetting printer that ejects ink droplets from a printhead. This printhead consists of a glass nozzle with a diameter of 50 µm and a piezoelectric transducer that is coated on the nozzle. The silver colloidal solution was inkjetted on a sintered CT800 ceramic substrate, followed by curing at 200 °C for 60 minutes. As a result, the silver trace with a thickness of 200 nm was obtained. The substrate, with the ejected silver thin film as the seed layer, was then immersed into a preinitiator solution to coat a layer of palladium for enhancing the deposition of nickel. Electroless nickel plating was successfully conducted at a rate of 0.39 µm /min, and the thickness of traces was plated up to 84 µm. This study demonstrates that the integration of inkjetting with plating is an effective method to form high aspect patterns at the demand location.

## I. Introduction

Micro inkjetting is an emerging technology that makes use of "additive drop on demand" technology. It can be used as a highly flexible manufacturing tool for patterning micro patterns including conductive traces and lines, passive circuits, dispensing and coating. Micro inkjetting can also be used as an alternative to expensive conventional patterning techniques such as photo-lithography. The advantages of micro inkjetting such as drop-on-demand and non-contact make it one of the most attractive, cost-effective and flexible manufacturing tools available today.

Due to the low viscosity of the ink and small diameter of the nozzle used in micro inkjet, the jetted ink only forms a thin film on substrates. Such thin traces not only cause high resistance in electrical connectivity but also result in incompatibility issues with soldering and wire bonding processes [2]. Hence, we aim to alleviate these limitations by integrating the inkjet process with electroless plating for realization of high aspect patterns.

The electroless nickel (EN) plating process is a popular maskless and selective deposition method used in many industries due to its ease of operation. EN also exhibits excellent material properties such as wear, corrosion resist-ance and solderability [3]. In this study, EN is used as a method for building up trace thickness.

This paper reports on the formation of high aspect patterns on sintered low temperature co-fired ceramic (LTCC) substrates by integrating micro inkjetting with electroless plating. Micro inkjetting was realized by using an inkjetting apparatus, which uses a piezoelectric transducer to drive a glass nozzle with a diameter of 50 µm to eject ink droplets. The silver trace with a line width of 70 µm and a thickness of 200 nm was obtained. The ejected silver thin film was used as the seed layer for electroless nickel (EN) plating. EN plating was then successfully conducted at a rate of 0.39 µm /min, and the thickness of traces was plated up to 84 µm. The achievements in this study demonstrated that integration of inkjetting with plating is an effective method to form high aspect patterns at the demand location.

## II. Experimental Methodology

### A. Substrate Material

The ceramic green tapes with a brand name of CT800 from Heraeus Germany were used as the starting material. The layer thickness of this tape was 127 µm; and five layers of this green tape were stacked and laminated by an isostatic laminator at 75 °C and 20 MPa for 10 minutes. The laminated substrate was placed in an oven for debinding and followed by a firing process in a furnace with a peak temperature of 850 °C. The sintered CT800 substrate was diced into pieces (20 mm× 20 mm) that were used for inkjet process. The following Table 1 illustrates the properties of CT800 ceramic material.

TABLE I
Properties of CT800 Ceramic Material

| Binder type | PVB |
|---|---|
| Diameter of ceramic particles | 2.2 µm |
| Green density | 1.78 g/cm3 |
| Tape thickness | 127 µm |
| Lamination parameters | 75°C/20 MPa/10 min |
| Sintering peak temperature | 850 °C |
| Sintered substrate thickness | 530 µm |
| Sintered surface roughness (Ra) | 0.63 µm |

### B. Process Flow for Integration

An inkjet printer (Jetlab® II) was used in this study. An orifice with 50 μm diameter was installed in the printhead for inkjet. Single-pass and multiple-pass inkjetting were carried out to understand the effect of surface roughness of the substrate to the inkjetted line width. The typical inkjetting parameters were set as: $T_r$ (rise time)= 3 μs, $T_d$ (dwell time)= 25 μs, $T_f$ (fall time)= 3 μs, V (voltage)= 30 V, S (printing speed) =10 mm/s and F (frequency) = 750 Hz.

The process flow was illustrated in Fig.1. The silver nano-particle ink was printed onto the substrate at room temperature, followed by curing at 200 °C for about 1 hour. This solvent-based silver ink is deposited as a seed layer for EN plating process. Prior to the actual electroless plating, the ceramic substrate was given a series of pre-treatment to remove any organic contamination and to activate the surface. The substrate was first immersed in a sulphuric acid solution for about 5 minutes (at 65 °C) to remove surface contamination such as oxides. The substrate was then immersed in an activating solution for 5 minutes (at 48°C) for nickel initiation, the purpose of which is to coat a thin palladium layer for activating the surface. After rinsing in deionized (DI) water, the substrate was immersed into the EN plating solution (at 85°C) for necessary duration to study the influencing factors of plating versus the conductor thickness and line width.

The apparatus used in the pre-treatments and electroless plating steps are similar. The setup consists of a glass beaker, a hot plate, a thermometer for temperature feedback and a magnetic stirrer for agitating the solution. Each plating step should be continuously carried out throughout the whole process in order to achieve consistent results.

The substrate was rinsed thoroughly with DI water and dried using nitrogen gas after nickel plating was completed. The width and thickness of plated profiles were characterized by an optical microscope and a stylus profilometer, respectively. The plated substrate was also sputtered with a thin layer of gold to facilitate energy dispersive X-ray analysis to study the elemental composition of the CT800 substrate.

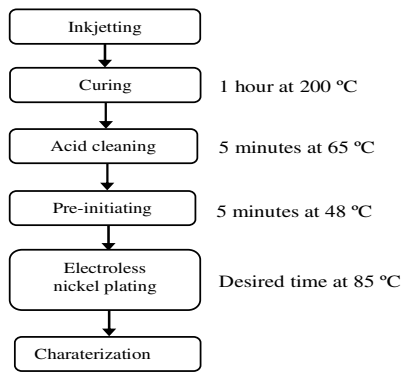

Fig.1. Process flow of integrating inkjetting with EN plating

### III. EXPERIMENTAL RESULTS

### A. Inkjetting on Ceramic Substrate

The surface roughness of the sintered CT800 substrate was about 0.63 μm (Ra). The free drop size was about 55 μm when a nozzle with 50 μm diameter was used for inkjet. However, the diameter of patterns inkjetted on the ceramic substrate varied from 60 μm to 100 μm due to the surface interaction between the ink and substrate. When the pitch on the substrate between two inkjet droplets changed from 70 μm to 100 μm, it was found that the droplets could not merge together to form continuous inkjet lines as the pitch was above 80 μm, and the optimum pitch was found to be about 70 μm. This pitch value was used in the subsequent experiments. Figs. 2 (a) and (b) shows optical images of inkjetted silver lines after (a) curing and (b) acid cleaning, respectively. The thickness of the single-pass inkjetted line on CT 800 could not be measured due to the roughness of the substrate surface; but this thickness was ranging from 0.2 μm to 0.4 μm on a glass surface. Multiple-pass lines were also studied to investigate the surface roughness of substrate and the effect onto EN plating. The surface roughness after multi-pass inkjetting was measured by a stylus profilometer, and as shown in Fig. 3, the surface roughness decreased with the increase of repeated inkjet passes since that the inkjetted nano particles smoothened the substrate surface.

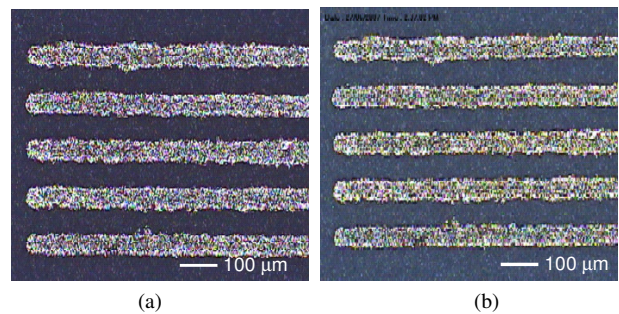

Fig. 2. Silver lines obtained via single-pass inkjetting. (a) After inkjetting and curing, (b) After acid cleaning

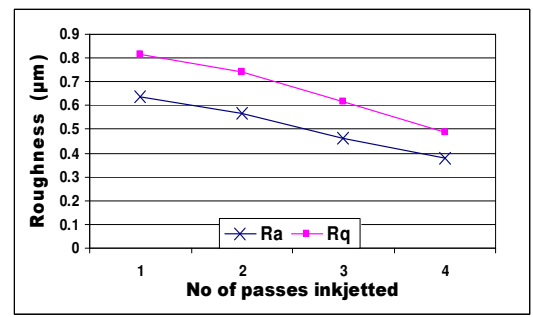

Fig. 3. Relationship between surface roughness of inkjetted pattern and numbers of inkjetted passes

*B. Electroless Nickel (EN) Plating*

After inkjetting and acid cleaning, the LTCC substrate with inkjetted silver traces was immersed into a preinitiating solution for a palladium coating. This coating ensures a complete coverage of the targeted patterns and will enhance metallic deposition during EN plating.

The entire electroless plating process should not be interrupted at any point of time during the process in order to achieve consistency. Fig. 4 shows the images of inkjetted silver lines after pre-initiating as in (a) and after 10 minutes of electroless plating as in (b).

*C. Line Width and Thickness versus Plating Time*

EN plating is not an anisotropic additive process. With the increase in plating thickness, the line width of targeted patterns increases as well. Fig. 5 (a) and (b) show the optical microscopy images of the line width after acid cleaning and after 120 minutes of EN plating. The increase in line width for this sample was found to be approximately 34 µm after 2 hours plating. Table 2 illustrates the increases of line width and thickness with the increase of plating duration.

The plated conductive traces showed a uniformly plated height under different plating durations, for instance, the average height of plated traces were about 12 µm after 30 minutes of EN plating. Fig. 6 shows the cross section of the line traces plated for different lengths of duration.

The thickness of the plated conductor traces was found to be nearly linearly proportional to the plating time at a rate of 21-24 µm/hr. The line width of the conductor traces had also increased with plating time. This was an expected result as the conductor traces were experiencing nickel deposition on the sidewalls as well, which resulted in growth in the x and y directions. Fig.7 shows a graphical summary of the results showing the plated height and variation in line width versus plating time.

A simple peel test was then performed on the plated conductive traces using Kapton tape. The adhesion was acceptable since no traces were peeled off in the peel test.

TABLE 2
THICKNESS AND WIDTH VERSUS PLATING TIME

| Measurement Parameter | Plating Time (minutes) | | | | |
|---|---|---|---|---|---|
| | 30 | 60 | 120 | 180 | 240 |
| Average plated height ( µm ) | 12 | 24 | 43 | 64 | 84 |
| Average change in line width ( µm ) | 2 | 14 | 34 | 73 | 100 |

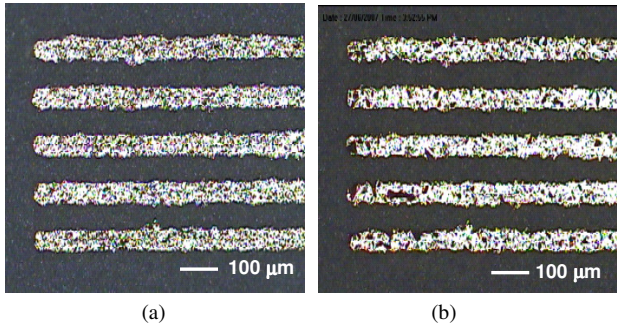

(a)                    (b)

Fig. 4. EN plating on inkjetted silver traces. (a) After preinitiating; (b) after 30 minutes plating.

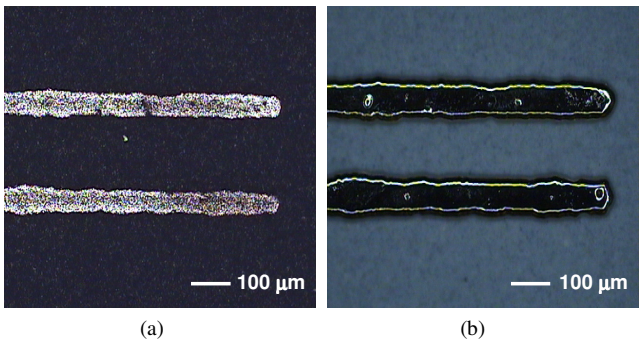

(a)                    (b)

Fig. 5. Optical images of line traces. (a) After cleaning but before plating, (b) after 120 minutes of EN plating.

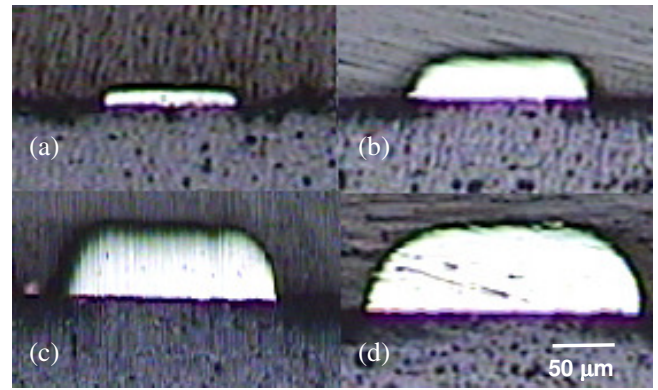

Fig. 6. Cross-section of conductive traces plated for (a) 30 mins; (b) 120 minutes; (c)180 minutes; (d) 240 minutes.

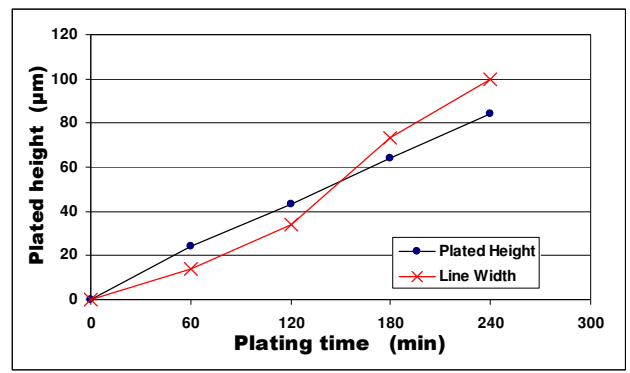

Fig. 7. The height and variation in line width versus plating time

### D. Effect of Inkjet Passes on Deposition Rates

The solid content of the silver ink used for deposition determines the metal content of seed layer after curing. The numbers of inkjet passes represent the repeated times of inkjetting at the same area. As the inkjet passes increase, the thickness of the seed layer would also increase. In this investigation, line traces of different numbers of passes (1-5 passes) were jetted onto the LTCC substrates to study the influence of trace passes on EN plating.

The plating rate of the line traces with different numbers of inkjet passes (1-5 passes) was found to be very similar. Fig.8 shows the scanned profiles of the line traces measured by a surface profilometer after EN plating. Fig. 9 shows the tabulated graphical results. These results also demonstrated that electroless plating was quite uniform for different pattern densities.

From the results in Fig. 8, the number of inkjet passes did not seem to have a significant influence on the EN plating rate. This was because that the EN deposition reaction took place at the interface of catalytic surface and the solution. Regardless of the numbers of inkjet passes, once there is a catalytic metal seed layer surface, EN deposition would take place provided that the chemical reactants in the solution were able to migrate to the interface. By precisely controlling the proper conditions for EN plating, a constant autocatalytic deposition reaction could take place and self propagate.

It is hence important that the EN solution be properly maintained and operated in a properly controlled environment to achieve optimum EN solution chemistry.

### E. Constrained Electroless Plating in Channels

In order to minimize the variation in line width and increase the aspect ratio of plated profiles, constrained plating process was proposed in this study. The silver seed layer was first inkjetted into the pre-patterned channels, followed by EN plating. The side walls of the patterned channels were used to constrain the lateral growth of nickel during plating. This resulted in the nickel being plated anisotropically in the vertical direction.

Micro channels were formed on laminated green ceramic substrates via micro embossing [7-9]. After embossing and sintering, it was found that micro channels with a depth of 43 μm and a width of 100 μm were formed on five-layered CT800 substrates. The inkjet nozzle and the channels were aligned via the visual system of the apparatus, and silver ink was deposited into the channels to seed a metal layer. Figs. 10 (a) and (b) show the optical images of the channels before and after plating.

Due to the deposition misalignment and flow behavior of the ink, there was a slight overflow of the silver ink across the channel. This has accounted for the resultant profile of the plated EN due to conformal plating of the silver seed layer as seen in Figs. 11 (a) and (b).

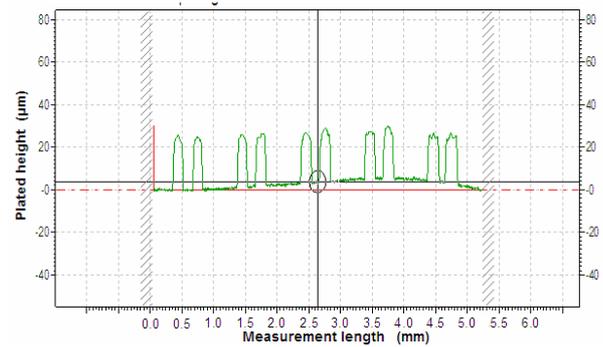

(a) After 60 minutes plating

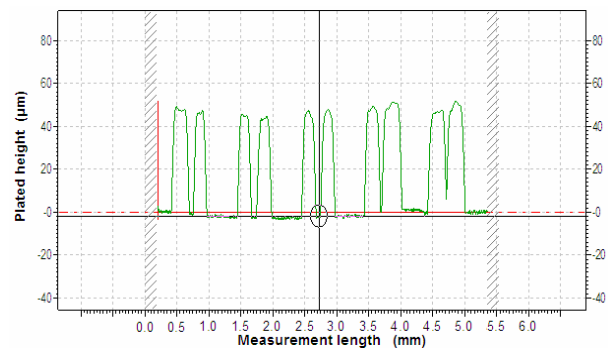

(b) After 120 minutes plating

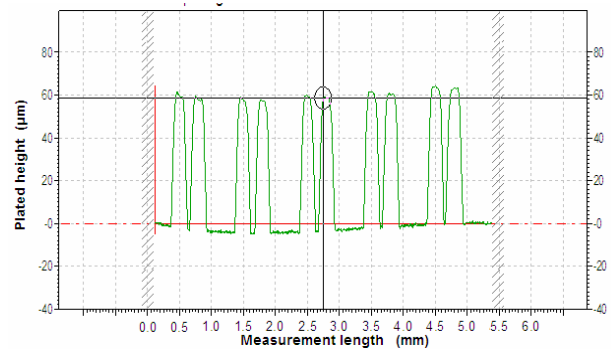

(c) After 180 minutes plating

Fig. 8. Scanned profile of traces with 1-5 (from left to right) passes subjected to (a) 60 minutes plating; (b) 120minutes plating; (c) 180 minutes plating.

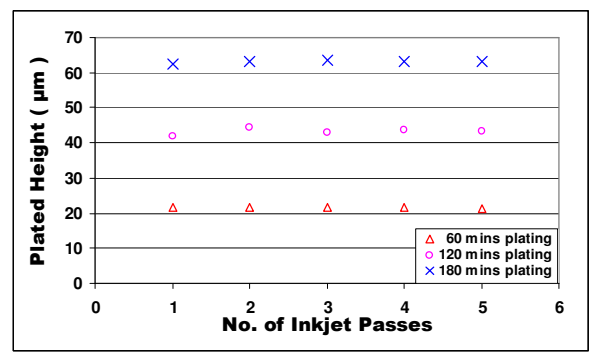

Fig. 9. Graphical results of inkjet passes on plating rate

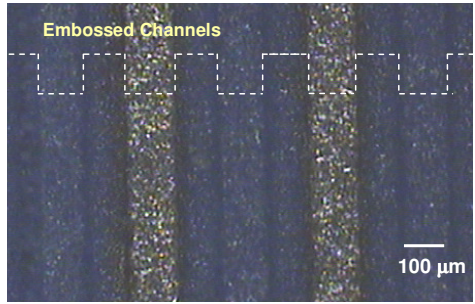

(a)    Channels before plating

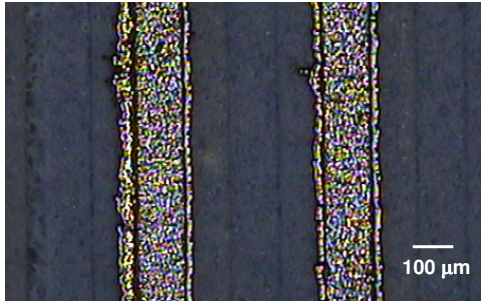

(b)    Channels after plating

Fig. 10. Optical images of channels. (a) Before plating; (b) After plating

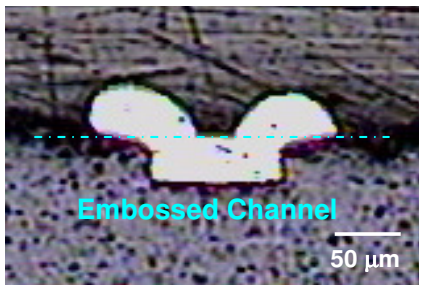

(a) 120 minutes of EN plating

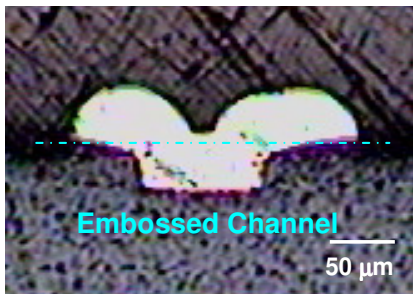

(b) 150 minutes of EN plating

Fig. 11. Cross section of EN plated channels. (a) 120 minutes plating;
(b) 150 minutes plating

Figs. 11 (a) and (b) show the cross-section of the plated substrates. These embossed channels were plated for 120 minutes and 150 minutes to achieve an approximate thickness of about 42 μm and 53 μm, respectively. The EN has completely filled the both embossed channels with great fidelity.   As the plating time increased from 120 minutes to 150 minutes, there was a significant change in the vertical growth of the traces. However, it was observed that the plated trace exhibited limited lateral growth within the 100 μm wide embossed channels, hence improved the feasibility of constrained electroless plating in channels.

*F.    Characterization of the Substrate with EDX*

After EN plating was performed, The CT800 substrate showed an obvious change in physical color. This color change possibly caused by the absorption of the plating solution or by the chemical changes that the substrates had undergone during plating.

In a preliminary study of how EN plating would affect the CT800 substrates, energy dispersive X-ray (EDX) was used to investigate the change in elemental composition of the substrates. In EDX analysis, the sampling area size was about 200 μm× 200 μm. Fig. 12 shows the EDX spectrum analysis of CT800 substrate prior to plating.

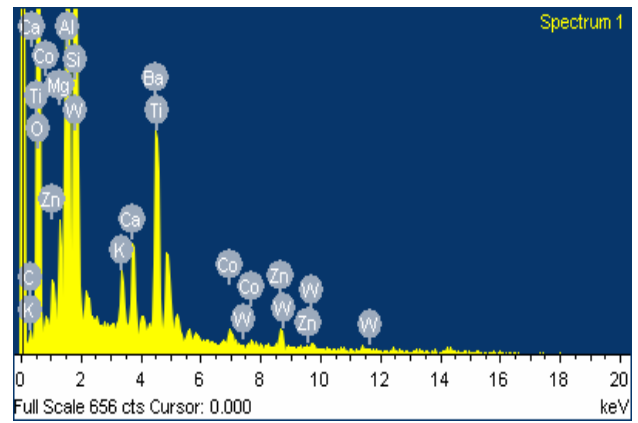

Fig. 12. EDX spectrum analysis of CT800 substrate before plating

It can be seen that the substrate was mainly composed of silicon, aluminum and oxygen, which were the common elements in LTCC materials. Other additive elements such as calcium, zinc, titanium and magnesium could also be found.

As the EN plating time increases, we noticed a decreasing trend in the atomic percentage of most elements except oxygen and aluminum. It was hypothesized that the compounds and trace metals in CT800 substrate have chemically reacted with the acidic EN plating solution, aluminum and oxygen, however, exist mainly as a very stable oxide in the form of $Al_2O_3$ and would be very chemically inert to the EN solution.

This initial EDX analysis revealed a change in chemical compositions between the plated and unplated CT800 substrates. Further characterization tests would be required and carried out to investigate the impact of electroless nickel plating on both mechanical and electrical properties of this ceramic substrates.

## IV. Conclusion

Formation of high aspect patterns via process integration of inkjetting and electroless plating on sintered LTCC substrates was studied. Both single pass and multiple passes of inkjet were conducted to study the effects of seed layer on the plating rate and had showed similar results. Electroless nickel plating was successfully conducted at a rate of 0.39 µm /min, and the thickness of traces was plated up to 84 µm. The plated conductive traces had not only increased in thickness but also exhibited a proportional increase in trace width, which led to limited aspect ratios of plated profiles.

Constrained plating by inkjetting the seed layers into the embossed channels was thus proposed as a solution to achieve high aspect ratio microstructures and promising results were obtained.

To conclude, the achievements demonstrated that process integration of inkjetting with plating is an effective method to form high aspect patterns. Further studies, including characterizations of solderability and bondability of the plated patterns will be reported in near future.

## Acknowledgment

The authors would like to express acknowledgments to Ms. HJ Lu, Dr. CW P Shi, Dr. YH Gan and Ms. HP Maw of SIMTech for their cooperation. This work is supported by Agency for Science, Technology and Research (A*STAR) of Singapore for Singapore-Poland Cooperation.